\documentstyle[12pt,fullpage,epsf]{article}
\begin{document}
\begin{center}
{\Large \bf Complete solutions of Altarelli-Parisi evolution equations in
leading order and structure functions at low-x}\\
\vspace{.75cm}
{\large R.Rajkhowa  and
J.K.Sarma\footnote{E-mail:jks@tezu.ernet.in}}\\
\vspace{.25cm}
Physics Department, Tezpur University, Napaam,\\
Tezpur-784 028, Assam, India \\
\end{center}
\vspace{.75cm}

\begin{abstract}
We present complete solutions of singlet and non-singlet Altarelli-Parisi (AP) evolution 
equations in leading order (LO) at low-x. We obtain t-evolutions of proton and neutron 
structure functions and x-evolutions of deuteron structure functions at low-x from AP 
evolution equations. The results of t-evolutions are compared with HERA low-x and low-$Q^{2}$ 
data and those of x-evolutions are compared with NMC low-x and low-$Q^{2}$ data. Also we compare 
our results with those of general solutions of AP evolution equations.
\end{abstract}
\vspace{.75cm}

\noindent {\large \bf 1. Introduction:}\\
\indent The Altarelli-Parisi (AP) evolution equations [1-4] are fundamental tools to study the 
t(=ln$(Q^{2}/\Lambda^{2})$)
and x-evolutions of structure functions, where x and $Q^{2}$ are Bjorken scaling and four momenta 
transfer in a deep inelastic scattering (DIS)  process [5] respectively and $\Lambda$ is the QCD cut-off
parameter. On the otherhand, the study of structure functions at low-x has become topical in view [6] of
high energy collider and supercollider experiments [7]. Solutions of AP evolution equations give quark and
gluon structure functions which produce ultimately proton, neutron and deuteron structure functions.
Though numerical solutions are available in the literature [8], the explorations of the possibility of 
obtaining analytical solutions of AP evoluation equations are always interesting. In this connection,
general solutions of AP evolution equations at low-x in leading order have already been obtained by applying  
Taylor expansion method [9] and t-evolutions [10] and x- evolutions [11] of structure functions have been 
presented.\\
\indent The present paper reports complete solutions of AP evolution equations in leading order at low-x and calculation
of t and x-evolutions for singlet and non-singlet structure functions and hence t- evolutions of proton and neutron structure 
functions and x-evolutions of deuteron structure functions. In calculating structure functions, input data points
have been taken from experimental data directly unlike the usual practice of using an input distribution function                  
introduced by hand. Results of proton and neutron structure functions are compared with the HERA low-x low-$Q^{2}$
data and those of deuteron structure functions are compared with the NMC low-x low-$Q^{2}$ data. Comparisons
are also made with the results of general solutions [10-12] of AP evolution equations
already obtained. In the present paper, section 1 is the introduction. In section 2 necessary theory has
been discussed. Section 3 gives results and discussions.\\
\vspace{.5cm}

\noindent{\large \bf 2. Theory:}\\
\indent Though the basic theory has been discussed elsewhere [10-12], the essential
steps of the theory has been presented here for clarity. The AP evolution equations
for singlet and non-singlet structure functions in the standard forms are [13]
\begin{eqnarray}
\frac{\partial F_{2}^{S}(x,t)}{\partial t}-\frac{A_{f}}{t}[\{3+4ln(1-x)\}
F_{2}^{S}(x,t)+ I_{1}^{S}(x,t)+I_{2}^{S}(x,t)]=0 
\end{eqnarray}
and
\begin{eqnarray}
\frac{\partial F_{2}^{NS}(x,t)}{\partial t}-\frac{A_{f}}{t}[\{3+4ln(1-x)\}
F_{2}^{NS}(x,t)+I^{NS}(x,t)]=0,
\end{eqnarray}
where,
\begin{eqnarray}
I_{1}^{S}=2\int_{x}^{1}\frac{dw}{1-w}\{(1+w^{2})F_{2}^{S}(\frac{x}{w},t)-2F_{2}^{S}(x,t)\},
\end{eqnarray}
\begin{eqnarray}
I_{2}^{S}=\frac{3}{2}N_{f}\int_{x}^{1}\{w^{2}+(1-w)^{2}\}G(\frac{x}{w},t)dw
\end{eqnarray}
and 
\begin{eqnarray}
I^{NS}=2\int_{x}^{1}\frac{dw}{1-w}\{(1+w^{2})F_{2}^{NS}(\frac{x}{w},t)-2F_{2}^{NS}(x,t)\}.
\end{eqnarray}
Here, $t=ln(Q^{2}/\Lambda^{2})$ and $A_{f}=4/(33-2N_{f})$, $N_{f}$ being the
number of flavours and $\Lambda$ is the QCD cut off parameter.\\ \\
\indent Using Taylor expansion method [9] and neglecting higher order terms as discussed
in our earlier works [10-12], $G(\frac{x}{w},t)$, $F_{2}^{S}(\frac{x}{w},t)$ and
$F_{2}^{NS}(\frac{x}{w},t)$ can be approximated for low-x as
\begin{eqnarray}
G\left(\frac{x}{w},t\right)\simeq G(x,t)+x\sum_{k=1}^{\infty}u^{k}\frac{\partial G(x,t)}{\partial x},
\end{eqnarray}
\begin{eqnarray}
F_{2}^{S}\left(\frac{x}{w},t\right)\simeq F_{2}^{S}(x,t)+x\sum_{k=1}^{\infty}u^{k}\frac{\partial F_{2}^{S}(x,t)}{\partial x}
\end{eqnarray}
and
\begin{eqnarray}
F_{2}^{NS}\left(\frac{x}{w},t\right)\simeq F_{2}^{NS}(x,t)+x\sum_{k=1}^{\infty}u^{k}\frac{\partial F_{2}^{NS}(x,t)}{\partial x}
\end{eqnarray}
where x=1-w.\\ \\

\indent Using equations (6), (7) and (8) in equations (3), (4) and (5) and performing u-integrations
we get
\begin{eqnarray}
I_{1}^{S}= -[(1-x)(x+3)]F_{2}^{S}(x,t)+[2xln(\frac{1}{x})+x(1-x^{2})]\frac{\partial F_{2}^{s}(x,t)}{\partial x},
\end{eqnarray}
\begin{eqnarray}
I_{2}^{S}= N_{f}[\frac{1}{2}(1-x)(2-x+2x^{2})G(x,t)+\{-\frac{1}{2}x(1-x)(5-4x+2x^{2})+\frac{3}{2}xln(\frac{1}{x})\}\frac{\partial G(x,t)}{\partial x}]
\end{eqnarray}
 and 
\begin{eqnarray}
I^{NS}= -[(1-x)(x+3)]F_{2}^{NS}(x,t)+[2xln(\frac{1}{x})+x(1-x^{2})]\frac{\partial F_{2}^{NS}(x,t)}{\partial t}.
\end{eqnarray}
Now using equations (9) and (10) in equation (1) we have,
\begin{eqnarray}
\frac{\partial F_{2}^{S}(x,t)}{\partial t}-\frac{A_{f}}{t}[A(x)F_{2}^{S}
(x,t)+B(x)\frac{\partial F_{2}^{S}(x,t)}{\partial x}+C(x)G(x,t)+D(x)\frac{\partial G(x,t)}{\partial x}]=0.
\end{eqnarray}
Let us assume for simplicity 
\begin{eqnarray}
G(x,t)=K(x)F_{2}^{S}(x,t),
\end{eqnarray}
where K(x) is a function of x. Our earlier assumption as more crude, K=constant. This assumption gives t-evolutions of
proton structure functions [8, 10, 12] which agree well with data. But it can not produce x-evolutions of deuteron
structure functions [11] which agree well with data, especially in high-$Q^{2}$ low-x region. Now equation (12) gives
\begin{eqnarray}
\frac{\partial F_{2}^{S}(x,t)}{\partial t}-\frac{A_{f}}{t}[L(x)F_{2}^{S}+M(x)\frac{\partial F_{2}^{S}(x,t)}{\partial x}]=0,
\end{eqnarray}
where,
$$A(x)=3+4ln(1-x)-(1-x)(3+x),$$
$$B(x)=x(1-x^{2})+2xln(\frac{1}{x}),$$
$$C(x)=N_{f}(1-x)(2-x+2x^{2}),$$
$$D(x)=-\frac{1}{2}N_{f}(1-x)(5-4x+2x^{2})+\frac{3}{2}ln(\frac{1}{x}),$$
$$L(x)=A(x)+K(x)C(x)+D(x)\frac{\partial K(x)}{\partial x}$$
and
$$M(x)=B(x)+K(x)D(x).$$
Secondly using equation (11) in equation (2) we have
\begin{eqnarray}
\frac{\partial F_{2}^{NS}(x,t)}{\partial t}-\frac{A_{f}}{t}[P(x,t)F_{2}^{NS}+Q(x,t)\frac{\partial F_{2}^{NS}(x,t)}{\partial x}]=0,
\end{eqnarray}
where, 
$$P(x)=3+4ln(1-x)-(1-x)(x+3)$$ 
and  
$$Q(x)=x(1-x^{2})-2xlnx.$$
The general solutions of equation (14) is
$$F(U,V)=0$$
where F is an arbitrary function and 
$$U(x,t,F_{2}^{S})=C_{1}$$
and
$$V(x,t,F_{2}^{S})=C_{2}$$
form a solution of equation
\begin{eqnarray}
\frac{dx}{A_{f}M(x)}=\frac{dt}{-t}=\frac{dF_{2}^{S}(x,t)}{-A_{f}L(x)F_{2}^{S}(x,t)}.
\end{eqnarray}
Solving equation (16) we obtain,
$$U(x,t,F_{2}^{S})=t\;exp[\frac{1}{A_{f}}\int\frac{1}{M(x)}dx]$$
and
$$V(x,t,F_{2}^{S})=F_{2}^{S}(x,t)\;exp[\int\frac{L(x)}{M(x)}dx].$$
If U and V are two independent solutions of equation (16) and if $\alpha$  and$\beta$ are arbitrary constants, then  
$$V=\alpha U+\beta$$ 
is called a complete solution of equation (16).\\ \\

\indent The complete solution [9]
$$F_{2}^{S}(x,t)\; exp[\int\frac{L(x)}{M(x)}dx]=\alpha t\;exp[\frac{1}{A_{f}}\int\frac{1}{M(x)}dx]+\beta$$
is a two-parameter family of planes. The one parameter family determined by taking $\beta=\alpha^{2}$ has equation 
\begin{eqnarray}
F_{2}^{S}(x,t)\; exp[\int\frac{L(x)}{M(x)}dx]=\alpha t\; exp[\frac{1}{A_{f}}\int\frac{1}{M(x)}dx]+\alpha^{2}.
\end{eqnarray}
Differentiating equation (17) with respect to $\alpha$, we get
$$\alpha =-\frac{1}{2}t\; exp[\frac{1}{A_{f}}\int\frac{1}{M(x)}dx].$$ 
Putting the value of $\alpha$ in equation (17), we obtain
\begin{eqnarray}
F_{2}^{S}(x,t)=-\frac{1}{4}t^{2}\; exp[\int(\frac{2}{A_{f}M(x)}-\frac{L(x)}{M(x)})dx].
\end{eqnarray}
Now, defining
$$F_{2}^{S}(x,t_{0})=-\frac{1}{4}t_{0}^{2}\; exp[\int(\frac{2}{A_{f}M(x)}-\frac{L(x)}{M(x)})dx],$$
at $t=t_{0}$, where $t_{0}=ln(Q_{0}^{2}/\Lambda^{2})$ at any lower value $Q=Q_{0}$, we get from equation (18)
\begin{eqnarray}
F_{2}^{S}(x,t)=F_{2}^{S}(x,t_{0})(\frac{t}{t_{0}})^{2}
\end{eqnarray}
which gives the t-evoluation of singlet structure function $F_{2}^{S}(x,t)$.\\ \\

\indent Proceeding exactly in the same way, and defining
$$F_{2}^{NS}(x,t_{0})=-\frac{1}{4}t_{0}^{2}\; exp[\int(\frac{2}{A_{f}Q(x)}-\frac{P(x)}{Q(x)})dx],$$
we get for non-singlet structure function 
\begin{eqnarray}
F_{2}^{NS}(x,t)=F_{2}^{NS}(x,t_{0})(\frac{t}{t_{0}})^{2},
\end{eqnarray}
which gives the t-evolution of non-singlet structure function $F_{2}^{NS}(x,t).$\\ \\

\indent Again defining,
$$F_{2}^{S}(x_{0},t)=-\frac{1}{4}t^{2}\; exp[\int(\frac{2}{A_{f}M(x)}-\frac{L(x)}{M(x)})dx]_{x=x_{0}},$$
we obtain from equation (18)
\begin{eqnarray}
F_{2}^{S}(x,t)=F_{2}^{S}(x_{0},t)\;exp[\int_{x_{0}}^{x}(\frac{2}{A_{f}M(x)}-\frac{L(x)}{M(x)})dx],
\end{eqnarray}
which gives the x-evoluation of singlet structure function $F_{2}^{S}(x,t)$. Similarly defining,
$$F_{2}^{NS}(x_{0},t)=-\frac{1}{4}t^{2}\; exp[\int(\frac{2}{A_{f}Q(x)}-\frac{P(x)}{Q(x)})dx]_{x=x_{0}},$$
we get
\begin{eqnarray}
F_{2}^{NS}(x,t)=F_{2}^{NS}(x_{0},t)\;exp[\int_{x_{0}}^{x}(\frac{2}{A_{f}Q(x)}-\frac{P(x)}{Q(x)})dx],
\end{eqnarray}
which determines the x-evoluation of non-singlet structure function $F_{2}^{NS}(x,t)$. Deutron, proton and neutron structure functions measured in deep inelastic electro-production can be written in terms of singlet and non-singlet quark distribution functions in leading order as
\begin{eqnarray}
F_{2}^{d}(x,t)=\frac{5}{9}F_{2}^{S}(x,t),
\end{eqnarray}
\begin{eqnarray}
F_{2}^{p}(x,t)=\frac{5}{18}F_{2}^{S}(x,t)+\frac{3}{18}F_{2}^{NS}(x,t)
\end{eqnarray}
and
\begin{eqnarray}
F_{2}^{n}(x,t)=\frac{5}{18}F_{2}^{S}(x,t)-\frac{3}{18}F_{2}^{NS}(x,t).
\end{eqnarray}
Now using equations (19) and (21) in equation (23) we will get t and x-evolution of deuteron structure function $F_{2}^{d}(x,t)$ at low-x as
\begin{eqnarray}
F_{2}^{d}(x,t)=F_{2}^{d}(x,t_{0})(\frac{t}{t_{0}})^{2}
\end{eqnarray}
and
\begin{eqnarray}
F_{2}^{d}(x,t)=F_{2}^{d}(x_{0},t)\;exp[\int_{x_{0}}^{x}(\frac{2}{A_{f}M(x)}-\frac{L(x)}{M(x)})dx],
\end{eqnarray}
where, the input functions are 
$$F_{2}^{d}(x,t_{0})=\frac{5}{9}F_{2}^{S}(x,t_{0})$$
and
$$F_{2}^{d}(x_{0},t)=\frac{5}{9}F_{2}^{S}(x_{0},t).$$
The corresponding results for general solutions of AP evolution equations obtained earlier [10-12] are 
\begin{eqnarray}
F_{2}^{d}(x,t)=F_{2}^{d}(x,t_{0})(\frac{t}{t_{0}})
\end{eqnarray}
and
\begin{eqnarray}
F_{2}^{d}(x,t)=F_{2}^{d}(x_{0},t)\;exp[\int_{x_{0}}^{x}(\frac{1}{A_{f}M(x)}-\frac{L(x)}{M(x)})dx].
\end{eqnarray}
These have been obtained by taking arbitrary linear combinations of U and V of arbitrary function $F(U,V)=0$ as solutions of equqations (14) and (15) for singlet 
and non-singlet structure functions respectively.\\ \\

\indent Similarly using equations (19) and (20) in equations (24) and 25) we get the t-evolutions of proton and neutron structure functions at low-x as
\begin{eqnarray}
F_{2}^{p}(x,t)=F_{2}^{p}(x,t_{0})(\frac{t}{t_{0}})^{2}
\end{eqnarray}
and
\begin{eqnarray}
F_{2}^{n}(x,t)=F_{2}^{n}(x,t_{0})(\frac{t}{t_{0}})^{2},
\end{eqnarray}
where, the input functions are 
$$F_{2}^{p}(x,t_{0})=\frac{5}{18}F_{2}^{S}(x,t_{0})+\frac{3}{18}F_{2}^{NS}(x,t_{0})$$
and
$$F_{2}^{n}(x,t_{0})=\frac{5}{18}F_{2}^{S}(x,t_{0})-\frac{3}{18}F_{2}^{NS}(x,t_{0}).$$
The corresponding results for general solutions of AP evolution equations are
\begin{eqnarray}
F_{2}^{p}(x,t)=F_{2}^{p}(x,t_{0})(\frac{t}{t_{0}})
\end{eqnarray}
and
\begin{eqnarray}
F_{2}^{n}(x,t)=F_{2}^{n}(x,t_{0})(\frac{t}{t_{0}}).
\end{eqnarray}
But the x-evoluations of proton and neutron structure functions like those of
deuteron structure function is not possible by this methodology, because to extract the x-evoluation of proton and neutron structure functions we are to put 
equations (21) and (22) in equations (24) and (25). But as the functions inside the integral sign of equations (21) and (22) are different, we need to separate 
the input functions $F_{2}^{S}(x_{0},t)$ and $F_{2}^{NS}(x_{0},t)$ from the data points to extract the x-evolutions of the proton and neutron structure functions, 
which is not possible.\\ \\
\vspace{.5cm}

\noindent{\large \bf 3. Results and Discussion:}\\
\indent In the present paper, we compare our results of t-evolutions of proton  and neutron
structure functions from equations (30) and (31) respectively with the HERA low-x,
low-$Q^{2}$ data [14]. Here proton structure functions $F_{2}^{p}(x,Q^{2},z)$
measured in the range $2\leq Q^{2}\leq 50\;GeV^{2}$, $0.73\leq z\leq 0.88$ and neutron structure
functions $F_{2}^{n}(x,Q^{2},z)$ measured in the range $2\leq Q^{2}\leq 50\;GeV^{2}$,
$0.3\leq z\leq 0.9$ have been used. Moreover here $P_{T}\leq 200\;Mev$, where $P_{T}$
is the transverse momentum of the final state baryon and  $z=1-q.(p-p')/(q.p)$, where
$p,q$ are the four momenta of the incident proton and the exchanged vector boson coupling
to the positron and $p'$ is the four-momentum of the final state baryon.\\ \\

\indent In figures 1(a-d) we present our results of t-evolutions of proton structure
functions $F_{2}^{p}$ (solid lines) for the representative values of x given in the
figures. Data points at lowest-$Q^{2}$ values in the figures are taken as input 
to test the evolution equation (30). Agreement is found to be excellent. In the same
figures we also plot the results of t-evolutions of proton structure functions $F_{2}^{p}$
(dashed lines) for the general solutions from equation (32) of AP evolution equations.
We observe that results of complete solutions are better than those of general solutions.\\ \\

\indent In figures 2(a-d) we present our results of t-evolutions of neutron structure
functions $F_{2}^{n}$ (solid lines) for the representative values of x given in the
figures. Data points at lowest-$Q^{2}$ values in the figures are taken as input 
to test the evolution equation (31). Agreement is found to be excellent. In the same
figures we also plot the results of t-evolutions of neutron structure functions $F_{2}^{n}$
(dashed lines) for the general solutions from equation (33) of AP evolution equations.
We observe that in this case also results of complete solutions are better than those of general solutions.\\ \\

\indent For a quantitative analysis of x-distributions of structure functions, we
evaluate the integrals that occured in equation (27) and present the results in figures
3(a-d) for $N_{f}=4$ and representative values of $Q^{2}$ given in each figure,
and compare them with NMC deuteron low-x low-$Q^{2}$ data [15]. In each figure the
data point for x-value just below 0.1 has been taken as input $F_{2}^{d}(x_{o},t)$.
If we take K(x)=constant, agreement of the results with experimental data becomes 
poor as before [11] especially in high-$Q^{2}$ lower-x region. We, therefore, consider here
K(x)=a.exp(-bx) where a=1 and b=1 for simplicity which gives better results. But explicit 
form of K(x) can actually be obtained only by solving coupled AP
evolution equations for singlet 
and gluon structure functions, and works are going on in this regard.

\indent Traditionally the AP equations provide a means of calculating the manner
in which the parton distributions change at fixed x as $Q^{2}$ varies. This change
comes about because of the various types of parton branching emission processes and
the x-distributions are modified as the initial momentum is shared among the
various daughter partons. However the exact rate of modifications of x-distributions at
fixed $Q^{2}$ can not be obtained from the AP equations since it depends not only
on the initial x but also on  the rate of change of parton distributions with respect to x,
$d^{n}F(x)/dx^{n}$ (n=1 to $\infty$), up to infinite order. Physically this implies that
at high-x, the parton has a large momentum fraction at its disposal and as a
result it radiates partons including gluons in innumerable ways, some of them involving complicated
QCD mechanisms. However for low-x, many of the radiation processes will cease to occur 
due to momentum constraints and the x-evoluations get simplified. It is then possible
to visualise a situation in which the modification of the x-distribution simply depends on
its initial value and its first derivative. In this simplified situation, the AP
equations give information on the shapes of the x-distribution as demonstrated in
this paper. The clearer testing of our results of x-evolution is actually the equation
(22) which is free from the additional assumption equation (13). But non-singlet
data is not sufficiently available in low-x to test our result. It is observed as a
whole that the results of complete solutions of AP evolution equations are improved
than those of general solutions, especially in t-evolution calculations. Of course, this
is a leading order calculation. Its natural improvement will be the calculation 
considering next-to-leading order terms and we plan to do so in our subsequent works.\\ \\ 
\vspace{.5cm}

\noindent{\large \bf Acknowledgement:}\\
\indent We are grateful to G. A. Ahmed of Department of Physics and A. K. Bhattacharya
of Department of Mathematical Sciences of Tezpur University for the help in numerical
part of this work. One of us (JKS) is grateful to UGC, New Delhi for the financial
assistance to this work in the form of a major research project.\\ 
\vspace{.5cm}

\noindent {\large \bf References:}\\
$[1]$G. Altarelli and G. Parisi, Nucl. Phys. B 126 (1977) 298.\\
$[2]$G. Altarelli, Phy. Rep. 81 (1981) 1.\\
$[3]$V.N. Gribov and L.N. Lipatov, Sov. J. Nucl. Phys. 20 (1975) 94.\\
$[4]$Y.L. Dokshitzer, Sov. Phy. JETP 46 (1977) 641.\\
$[5]$Francis Halzen and  Alan D. Martin, Quarks and Leptons: An Introductory
   Course in Modern Particle Physics, John and Wiley (1984).\\
$[6]$See for example, A.Ali and J.Bartels, eds. Proc. DESY-Topical Meeting on the
    Small-x Behaviour of Deep Inelastic Structure Functions in QCD, 1990, North-Holland,
            Amsterdam, (1991).\\
$[7]$See for example, G. Jariskog and D. Rein, eds. Proc. Large Hardon Collider Workshop,
      CERN report, CERN 90-10, (1990).\\
$[8]$D.K. Choudhury and J.K Sarma, Pramana-J. Phys.  39 (1992) 273.\\
$[9]$Frank Ayres Jr., Differential Equations, Schaum's Outline Series, McGraw-Hill (1952).\\
$[10]$J.K. Sarma and B. Das, Phy. Lett. B304 (1993) 323.\\
$[11]$J.K. Sarma, D.K. Choudhury and G.K. Medhi, Phys. Lett. B403(1997)139.\\
$[12]$D.K. Choudhury and J.K. Sarma, Pramana-J. Phys. 38 (1992) 481.\\
$[13]$L.F. Abbott, W.B. Atwood and R.M. Barnett, Phys. Rev. D22 (1980) 582.\\
$[14]$C. Adloff et al., H1 Collaboration, DESY 98-169, hep-ex/9811013, (1998).\\
$[15]$M. Arneodo et al., hep/961031, NMC, Nucl. Phy. B483 (1997).\\
\vspace{.5cm}

\begin{figure}[htbp]
    \begin{center}
    \leavevmode
    \centerline{\epsfbox{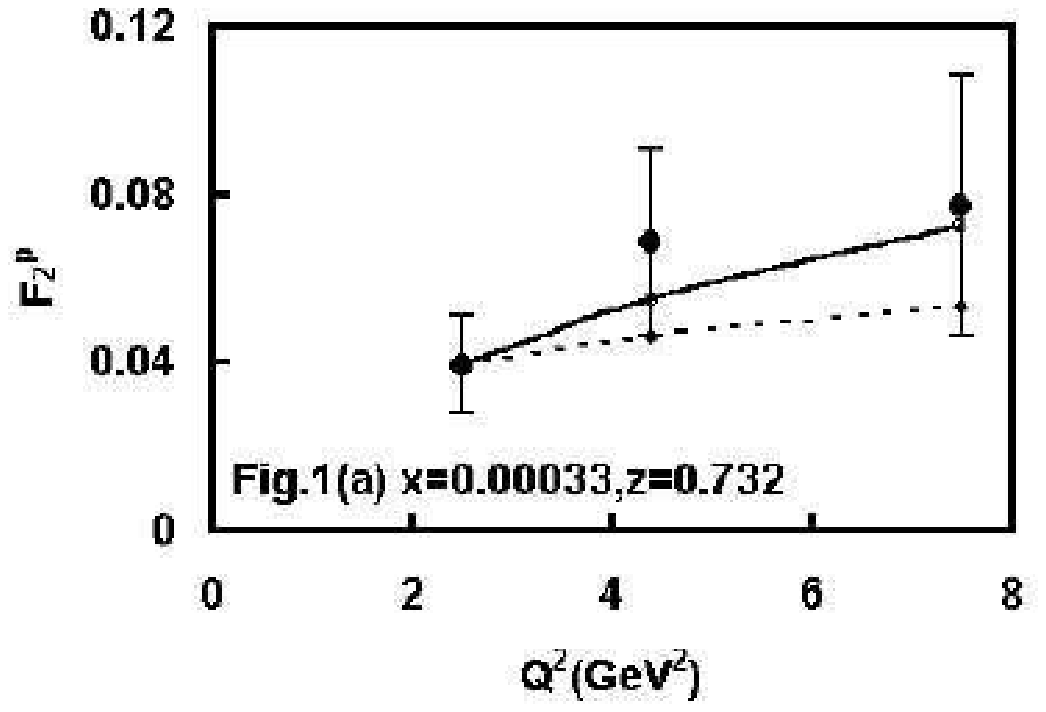}}
    \end{center}
\end{figure}

\begin{figure}[htbp]
    \begin{center}
    \leavevmode
    \centerline{\epsfbox{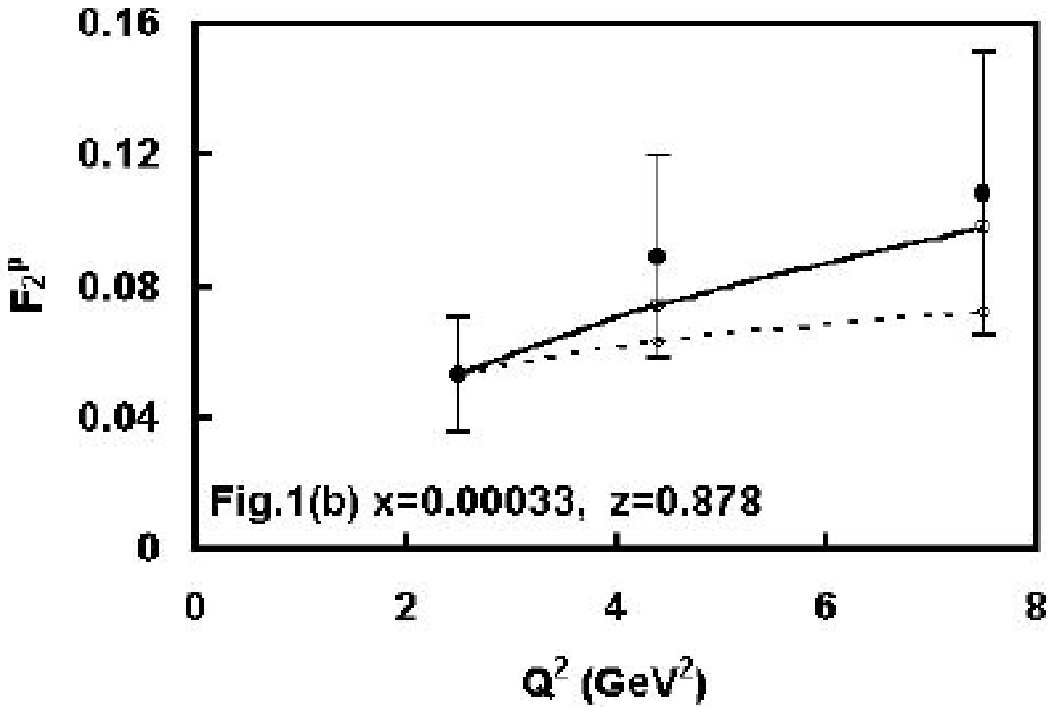}}
    \end{center}
\end{figure}

\begin{figure}[htbp]
    \begin{center}
    \leavevmode
    \centerline{\epsfbox{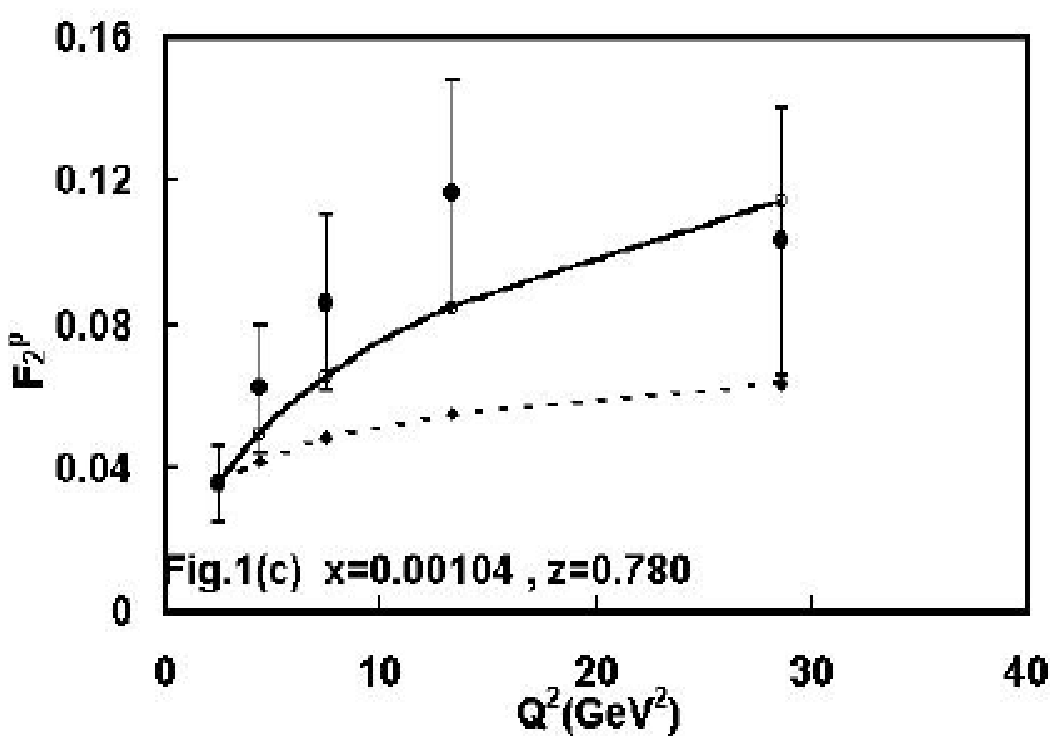}}
    \end{center}
\end{figure}

\begin{figure}[htbp]
    \begin{center}
    \leavevmode
    \centerline{\epsfbox{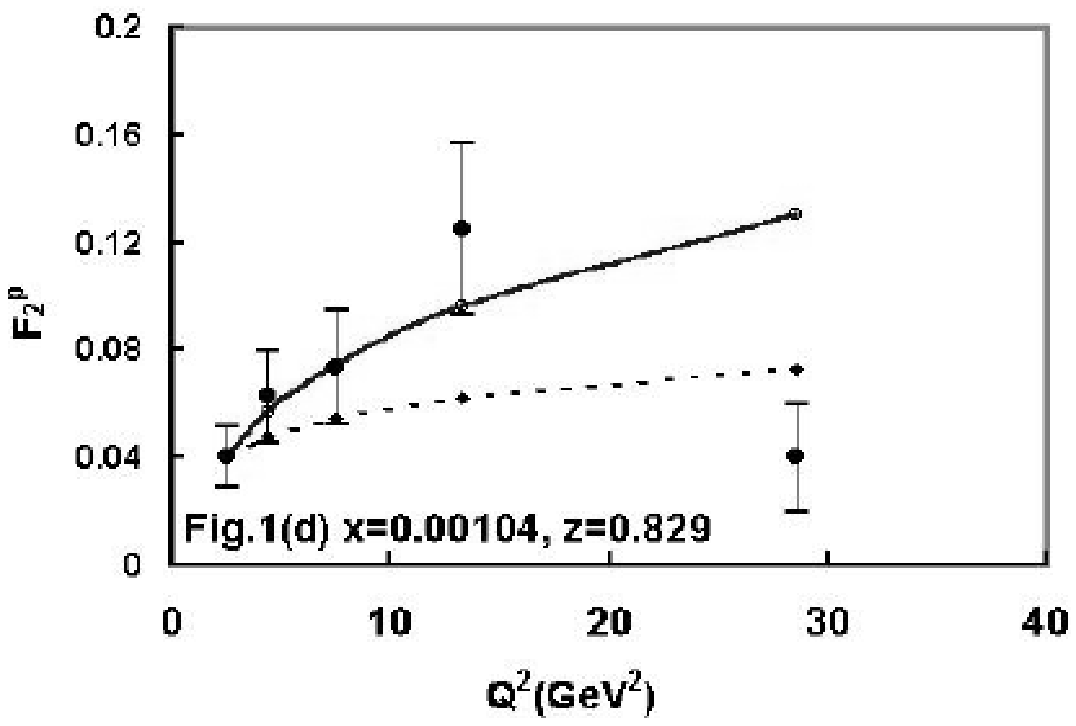}}
    \end{center}
\caption {In figures 1(a-d) t-evolutions of proton structure functions $F_{2}^{p}$
(solid lines) for the representative values of x given in the figures. Data points
at lowest-$Q^{2}$ values in the figures are taken as input to test the evolution equation (30).
To test the evolution equation (30). In the same figures we also plot the results of
t-evolutions of proton structure functions $F_{2}^{p}$ (dashed lines) for the
general solutions from equation (32) of AP evolution equations.}
\end{figure}

\begin{figure}[htbp]
    \begin{center}
    \leavevmode
    \centerline{\epsfbox{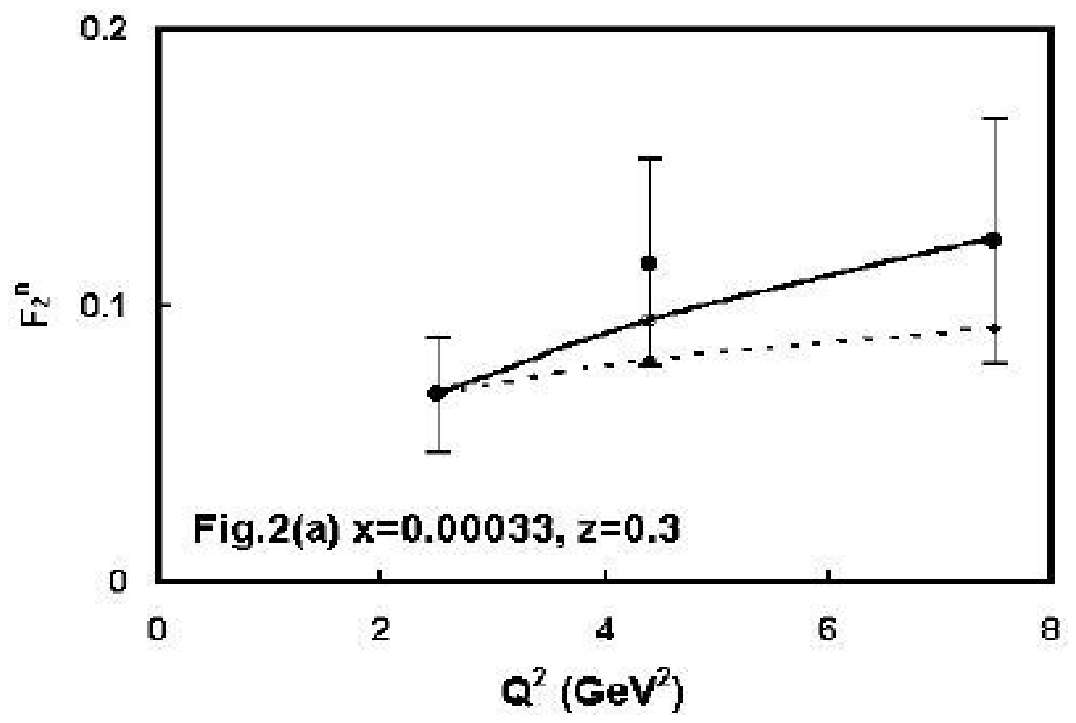}}
    \end{center}
\end{figure}

\begin{figure}[htbp]
    \begin{center}
    \leavevmode
    \centerline{\epsfbox{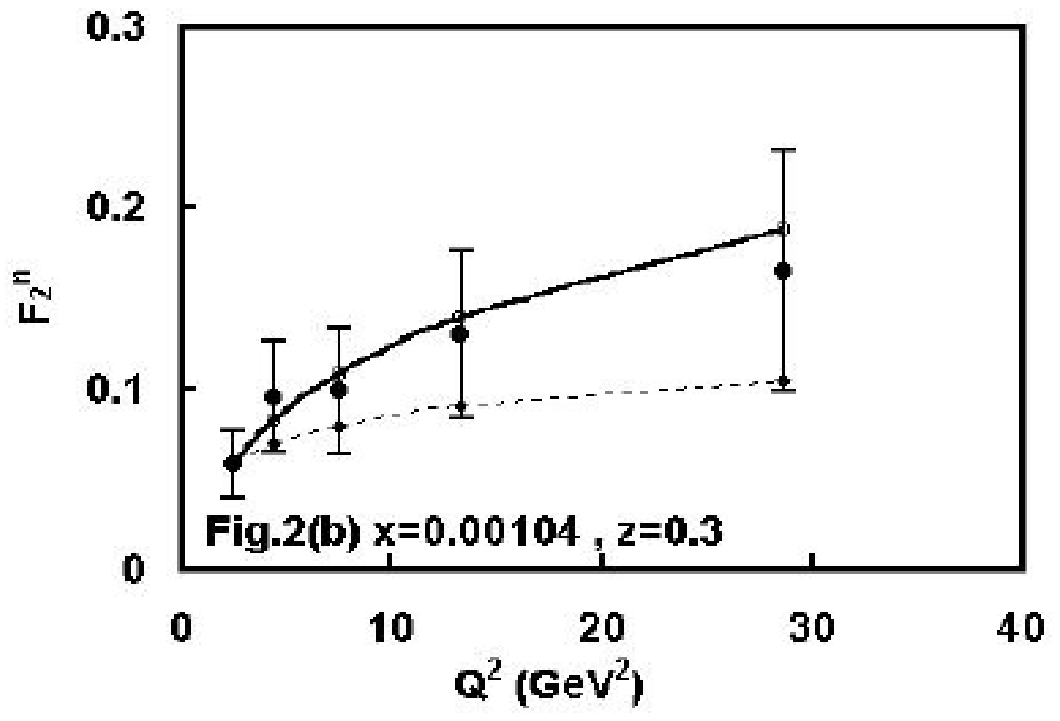}}
    \end{center}
\end{figure}

\begin{figure}[htbp]
    \begin{center}
    \leavevmode
    \centerline{\epsfbox{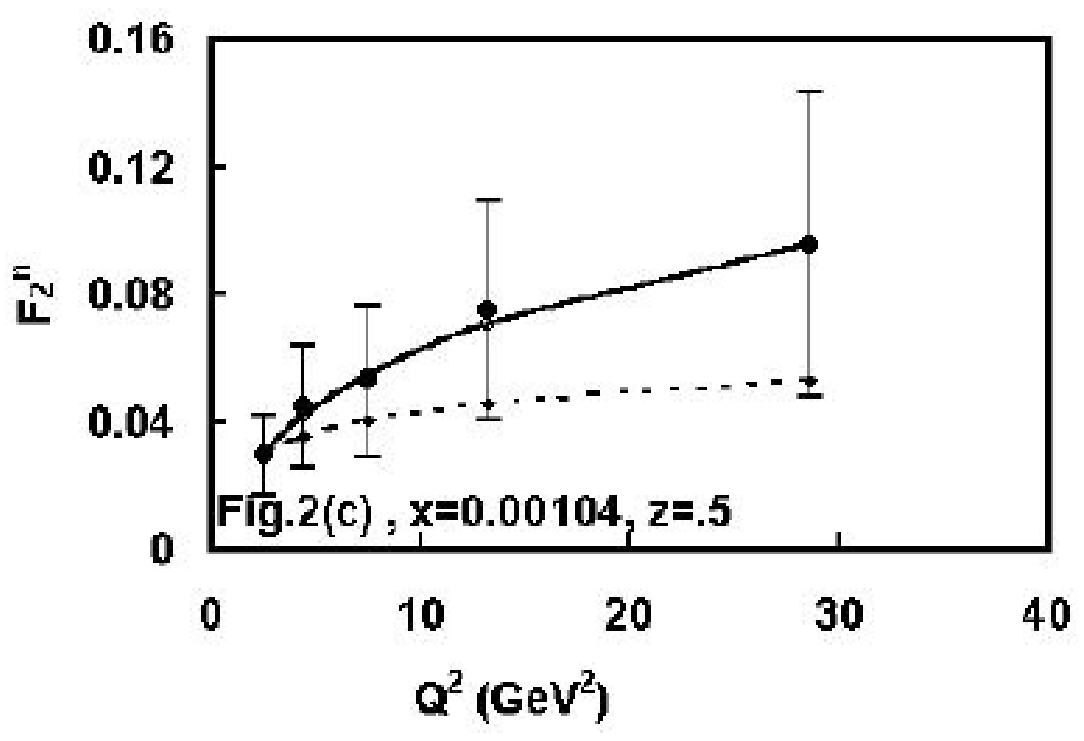}}
    \end{center}
\end{figure}

\begin{figure}[htbp]
    \begin{center}
    \leavevmode
    \centerline{\epsfbox{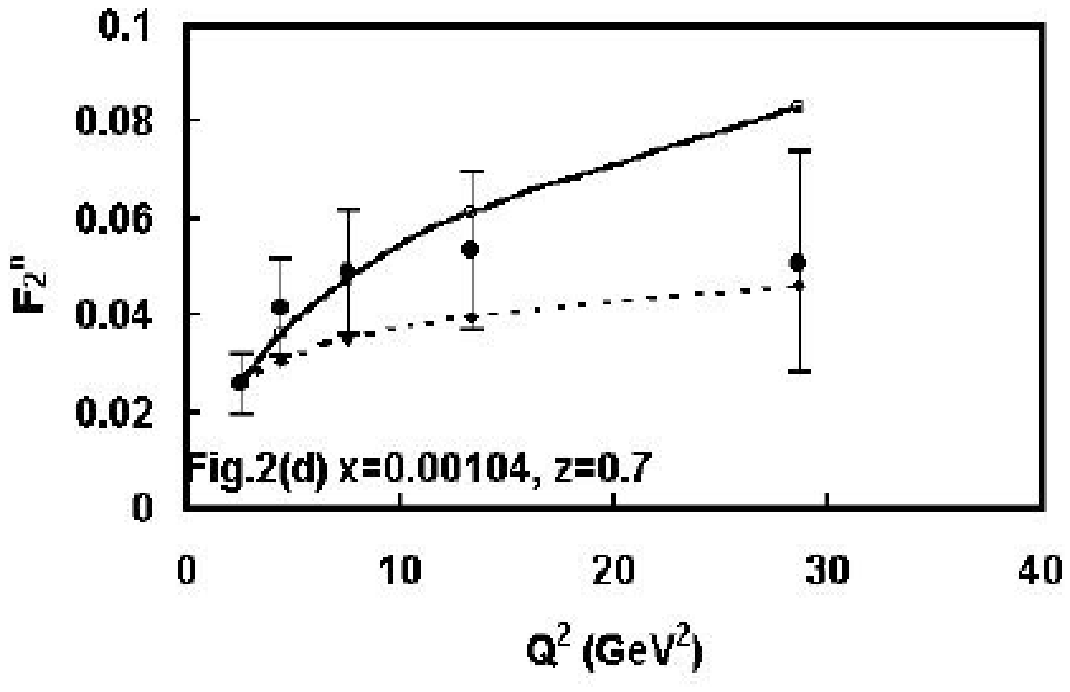}}
    \end{center}

\caption {In figures 2(a-d) t-evolutions of neutron structure
functions $F_{2}^{n}$ (solid lines) for the representative values of x given in the figures. Data
points at lowest-$Q^{2}$ values in the figures are taken as input to test the evolution equation (31).
In the same figures we also plot the results of t-evolutions of neutron structure
functions $F_{2}^{n}$ (dashed lines) for the general solutions from equation (33) of
AP evolution equations.}
\end{figure}

\begin{figure}[htbp]
    \begin{center}
    \leavevmode
    \centerline{\epsfbox{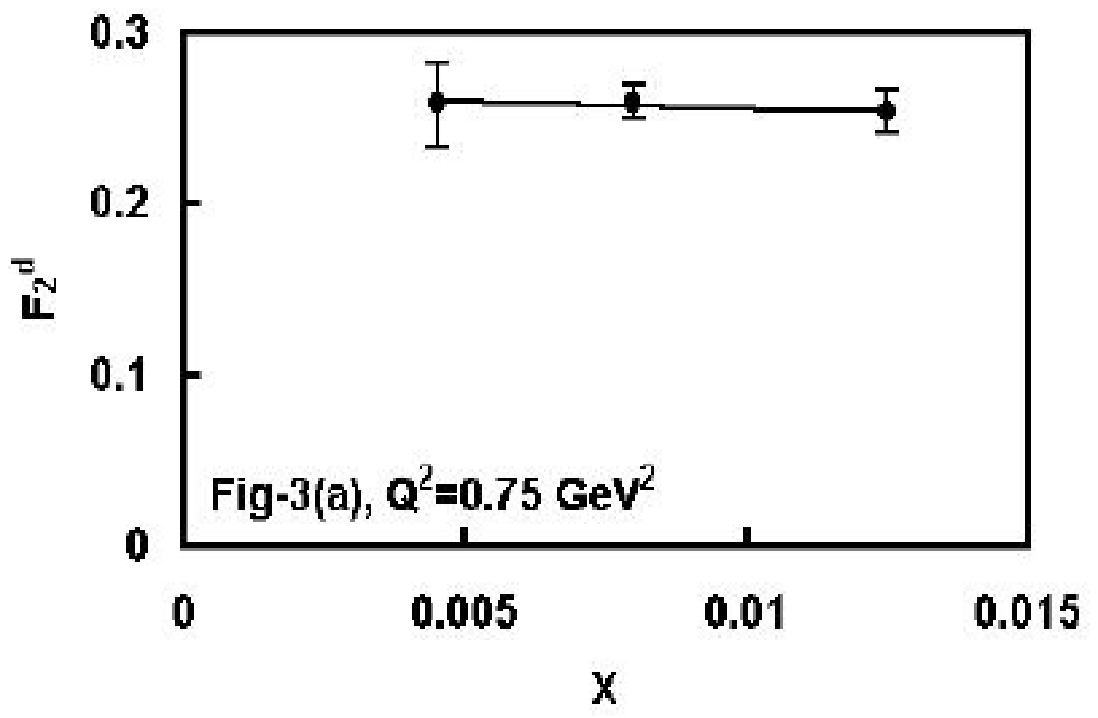}}
    \end{center}
\end{figure}

\begin{figure}[htbp]
    \begin{center}
    \leavevmode
    \centerline{\epsfbox{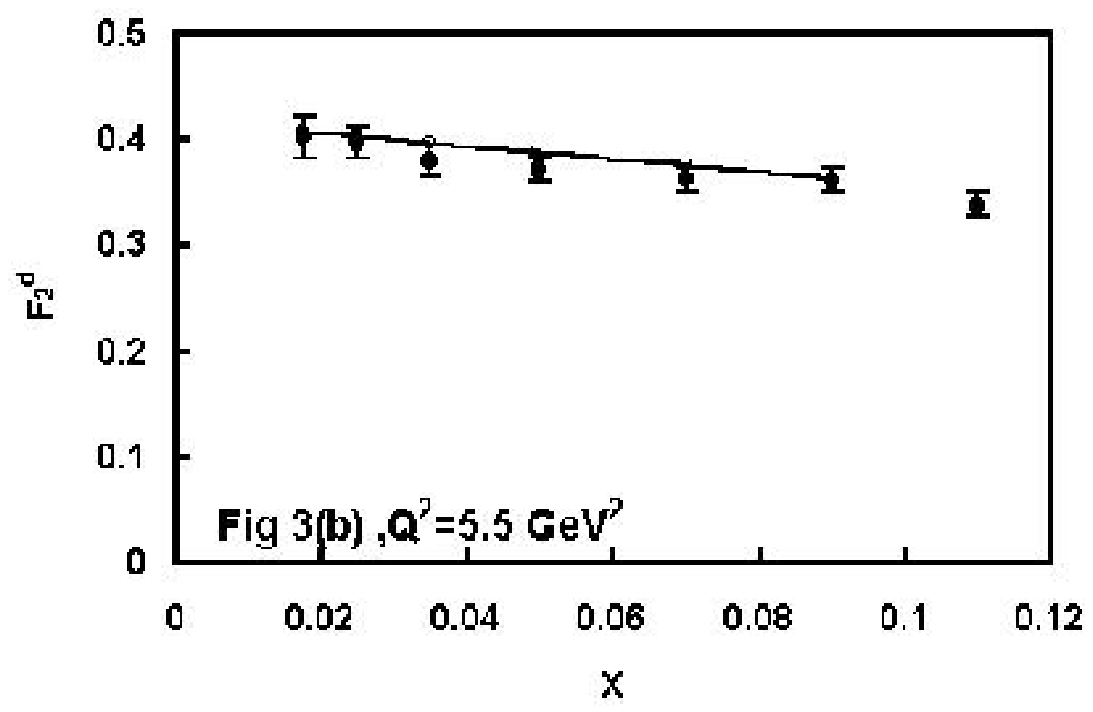}}
    \end{center}
\end{figure}

\begin{figure}[htbp]
    \begin{center}
    \leavevmode
    \centerline{\epsfbox{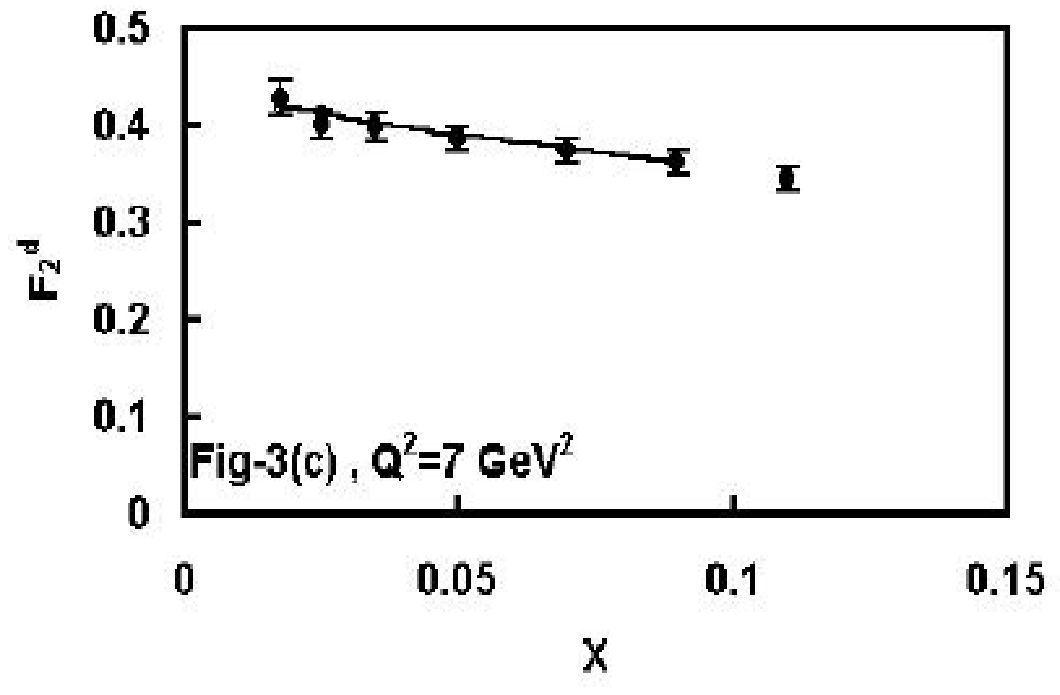}}
    \end{center}
\end{figure}

\begin{figure}[htbp]
    \begin{center}
    \leavevmode
    \centerline{\epsfbox{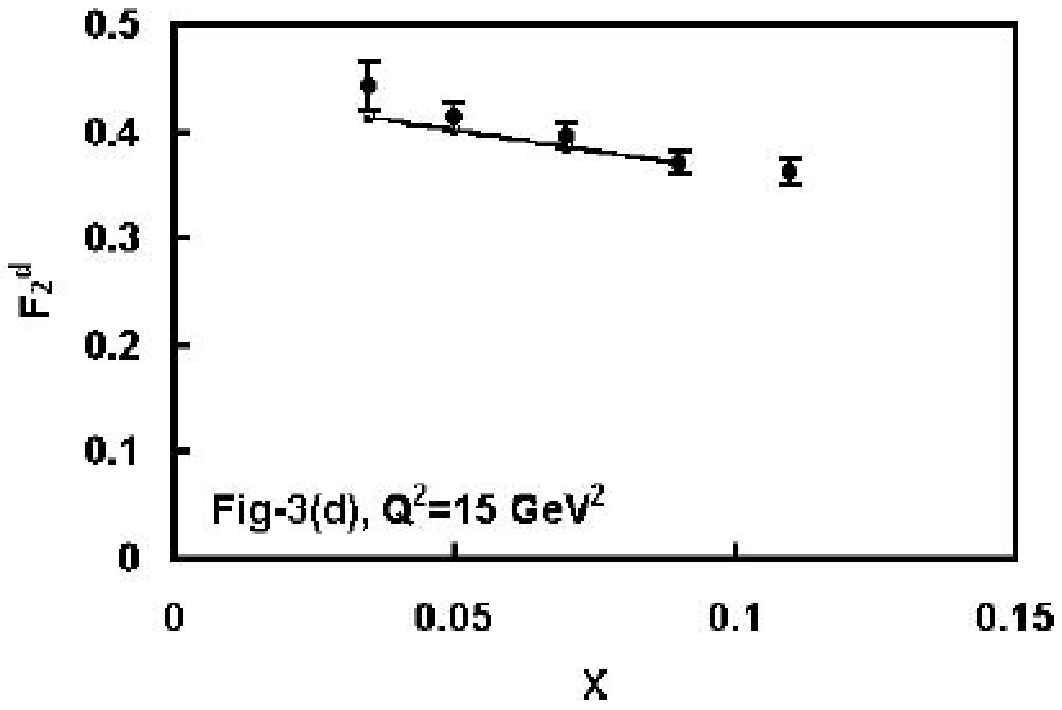}}
    \end{center}

\caption {In figures 3(a-d) x-distributions of
deuteron structure functions (solid lins) for $N_{f}=4$ and the representative values of $Q^{2}$
given in each figure and compared them with NMC deuteron low-x low $Q^{2}$ data.
In each figure the  data point for x-value just below 0.1 has been taken as
input $F_{2}^{d}(x_{0},{t}).$}
\end{figure}
\end{document}